\documentclass[a4paper]{jpconf}
\usepackage{graphicx}
\usepackage{epsfig}
\newcommand{\GERDA}{\textsc{Gerda}}
\newcommand{\gerda}{\textsc{Gerda}}

\newcommand{\ROOT}{\textsc{ROOT}}
\newcommand{\Majorana}{\textsc{Majorana}}
\newcommand{\MGDO}{\textsc{MGDO}}
\newcommand\Cpp{C{}\texttt{++}}

\begin{document}
\title{The MGDO software library for data analysis in 
Ge neutrinoless double-beta decay experiments}

\author{M.~Agostini$^{1}$, J.~A.~Detwiler$^{2}$, P.~Finnerty$^{3}$, 
K.~Kr\"oninger$^{4,a}$, D.~Lenz$^{4,b}$, J.~Liu$^{4,c}$, M.~G.~Marino$^{5,d}$, 
R.~Martin$^{2}$, K.~D.~Nguyen$^{2}$, L.~Pandola$^{6}$, 
A.~G.~Schubert$^{5}$, O.~Volynets$^{4}$ and P.~Zavarise$^{6,7}$}
\address{$^{1}$ Technische Universit\"at M\"unchen (TUM), Munich, Germany}
\address{$^{2}$ Lawrence Berkeley National Laboratory, Berkeley (CA), United States}
\address{$^{3}$ University of North Carolina, Chapel Hill (NC), United States}
\address{$^{4}$ Max-Planck-Institut f\"ur Physik, Munich, Germany}
\address{$^{5}$ University of Washington, Seattle (WA), United States}
\address{$^{6}$ INFN, Laboratori Nazionali del Gran Sasso, Assergi, Italy}
\address{$^{7}$ University of L'Aquila, L'Aquila, Italy}
\address{$^{a}$ Presently at University of G\"ottingen, G\"ottingen, Germany}
\address{$^{b}$ Presently at University of Wisconsin, Madison (WI), United States}
\address{$^{c}$ Presently at Institute for the Physics and Mathematics of the Universe, 
Tokyo, Japan}
\address{$^{d}$ Presently at Technische Universit\"at M\"unchen (TUM), Munich, Germany}

\ead{JADetwiler@lbl.gov, pandola@lngs.infn.it, volynets@mppmu.mpg.de}

\begin{abstract}
The \gerda\ and \Majorana\ experiments will search for neutrinoless double-beta decay of 
$^{76}$Ge using isotopically enriched high-purity germanium detectors.
Although the experiments differ in conceptual design, they share many aspects in common, 
and in particular will employ similar data analysis techniques. 
The collaborations are jointly 
developing a \Cpp\ software library, \MGDO, which contains a set of data objects and interfaces to encapsulate, store 
and manage physical quantities of interest, such as waveforms and high-purity germanium detector geometries. 
These data objects define a common format for persistent data, whether it is generated by
Monte Carlo simulations or an experimental apparatus, to reduce code duplication and to
ease the exchange of information between detector systems.
\MGDO\ also includes general-purpose analysis tools that can be used for the processing of 
measured or simulated digital signals.
The \MGDO\ design is based on the Object-Oriented programming paradigm 
and is very flexible, allowing for easy extension 
and customization of the components.
The tools provided by the \MGDO\ libraries are 
used by both \GERDA\ and \Majorana.
\end{abstract}

\section{Introduction and requirements}
The \gerda ~\cite{gerda} and \Majorana ~\cite{majorana} experiments are designed to 
search for the neutrinoless double-beta ($\beta \beta$) decay of $^{76}$Ge using 
isotopically-enriched high-purity germanium 
(HPGe) detectors. The experiments differ in basic 
approach and in the shielding philosophy for background suppression. 
Detectors in \gerda\ 
are directly immersed in liquid argon, which acts both as cooling 
medium and as passive shielding against external $\gamma$ radiation. 
The commissioning of \gerda\ was successfully completed and the Phase~I data 
taking has recently started with  
eight HPGe detectors enriched in $^{76}$Ge (total mass of about 15~kg). 
The design adopted 
by \Majorana\ consists in operating HPGe detectors in an ultra-radiopure vacuum 
cryostat surrounded by passive lead shielding.
The \Majorana\ 
Collaboration is currently constructing two modules of HPGe detectors, 
with a total mass of about 40~kg of natural and enriched germanium, and is scheduled to begin data taking with the first module in 2013.

In spite of the different shielding design of the two experiments, they use 
many of the same materials and employ similar detector technologies.
Since 2004, \gerda\ and \Majorana\ have taken advantage of this overlap 
by collaborating in an  
open way on topics of common interest. In particular they jointly develop and maintain
 a common and flexible 
Monte Carlo framework, \textsc{MaGe}~\cite{mage}. 
The advantage of having a shared software project is 
that it avoids duplication of effort for software tools of common interest, thus easing 
debugging and validation efforts. 
Following this experience, a new shared software project has been initiated by  
\gerda\ and \Majorana\, called  \Majorana -\gerda\ Data Objects (\MGDO). 
The core function of this software is to provide a collection of \Cpp\ objects to 
encapsulate HPGe detector array event data and related analytical quantities.
\MGDO\ thus provides a shared standard
for the storage, access, and manipulation of \gerda\ and \Majorana\ data. It also includes 
implementations of a number of general-purpose signal processing algorithms
to support advanced detector signal analysis.

\section{Basic concept and design}
\MGDO\ is designed as a set of class libraries exploiting 
the Object-Oriented paradigm. Virtual base classes are provided which define abstract interfaces.
Such an approach makes the code flexible, easing future extensions and user customization. 
The source code is written in the \Cpp\ programming language, allowing for the interfacing and extension of \MGDO\ with other general-purpose software for scientific computing. In particular, \MGDO\ adopts the system of units, physical constants, 
and several other tools provided by the CLHEP 
libraries~\cite{clhep}. \MGDO\ may also optionally be built against \textsc{FFTW3}~\cite{fftw3} 
for its fast digital Fourier-transform routines, 
and may be built against \ROOT~\cite{root} for its advanced data storage and analysis tools.

The tools provided in \MGDO\ can be divided in two main categories: data objects and transforms. The ``data objects''
are \Cpp\ classes which encapsulate complex data and physical quantities, for example digitized 
waveforms, detector parameters, and channel cross-talk matrices. 
The aim of the data objects is to support data management and handling: the existence of an  
\MGDO-defined standardized format eases the exchange and the cross-comparison of information, for example between Monte Carlo simulations and experimental data, or between different detector systems. \ROOT\ wrappers for the data objects are also provided, 
which makes possibles the storage of \MGDO\ data objects in a \ROOT-based format. A concrete example of one of our key data objects, MGWaveform, is presented in section~\ref{sec3}.

The other main category of tools provided by \MGDO\ are ``transforms'', general-purpose 
algorithms for digital signal processing. They encompass digital filters and other 
utilities, such as as smoothing, 
differentiation and extremum finders. Fourier and wavelet transform routines are provided to
enable analysis in the time domain, the frequency domain, or both. 
Having many commonly-used algorithms 
available avoids code duplication in experiment-specific analysis software, and thus 
enhances testing, validation and performance optimization. 
\ROOT\ wrappers for transforms are also provided to enable their use in interactive \ROOT\ sessions.  
Transforms are discussed in more detail in section~\ref{sec4}.

\section{A concrete data object: MGWaveform} \label{sec3}
One of the key data objects provided by \MGDO\ is MGWaveform.
It is designed to store and manipulate a digitized waveform produced by any kind of detector 
(an HPGe detector, a photo-multiplier, etc.). 
At its heart, MGWaveform is simply an array
containing the individual samples of the waveform, plus a number of
auxiliary attributes such 
as the sampling frequency, a time value to associate with the start of the trace, and 
the waveform type (e.g.~voltage pulse, current pulse, etc.). 
The public interface of the class consists mostly of   
protected Get and Set methods for the attribute I/O,
and of custom operators implementing  
widely-used waveform-waveform or waveform-scalar operations (product, sum, etc.). 
The definition of 
custom operators facilitates the basic handling of waveform
by internally managing the loop over the 
individual samples and by providing a consistent output. 

The ``\ROOT ified'' version of the class, MGTWaveform,
has additional public methods which are inherited from the \ROOT\ interface and provide  
direct connection to the \ROOT\ I/O, graphical, and analytic utilities.
For example, MGTWaveform provides functions to represent the waveform as
a histogram object (TH1) that can be drawn or fit. It also provides methods to
set a waveform from a function object (TF1), or to treat the waveform as a TF1 to be used
itself as a fit function.

An advantage of MGWaveform is that it can be used as a standard output format. 
Experimental 
waveforms can be stored as MGWaveform objects, while the output of Monte Carlo 
simulations can be also directly produced as MGWaveforms. This allows the real and simulated waveforms 
to be treated on exactly the same footing by the analysis, easing inter-comparison, validation, 
adoption of simulated pulses into analysis routines, etc.
Furthermore, having a common format facilitates the exchange of information 
between \gerda\ and \Majorana, including data and Monte Carlo results of common interest. 

Other data objects are available in \MGDO, as for instance MGTEvent. MGTEvent is an \MGDO\ class 
designed to encapsulate the full information of events: it includes a set of 
waveforms (one per channel) and additional data attributes, such as a time stamp and DAQ flags.

\section{The transforms} \label{sec4}
The \MGDO\ transforms are general-purpose algorithms for operations 
on MGWaveform data objects. The possible output of a transform is either a new MGWaveform 
(e.g.~for filters or smoothers), or a set of scalar parameters (e.g.~rise times or extrema), or both. Each transform is implemented  
as a class inherited from the general virtual class MGVWaveformTransformer. The base class provides a general 
interface making it easy to add new user-specific transforms. 

Implemented transforms that give a new waveform as output include ones which calculate and subtract a baseline, 
calculate the numeric derivative using simple (three-point) or more complex (five-point or RC derivative) 
algorithms, smooth the waveform using a moving average or triangular smoothing, and apply a trapezoidal 
filter \cite{TrapezFilter}.
Some of the scalar parameters calculated by transforms based on the input waveform(s) include the $\chi^2$ difference of 
two waveforms, the global maximum or minimum of a waveform, and the rise time within given limits, e.g.~10\% and 90\% of the amplitude.

\begin{figure}[h]
  \begin{center}

  \epsfig{file=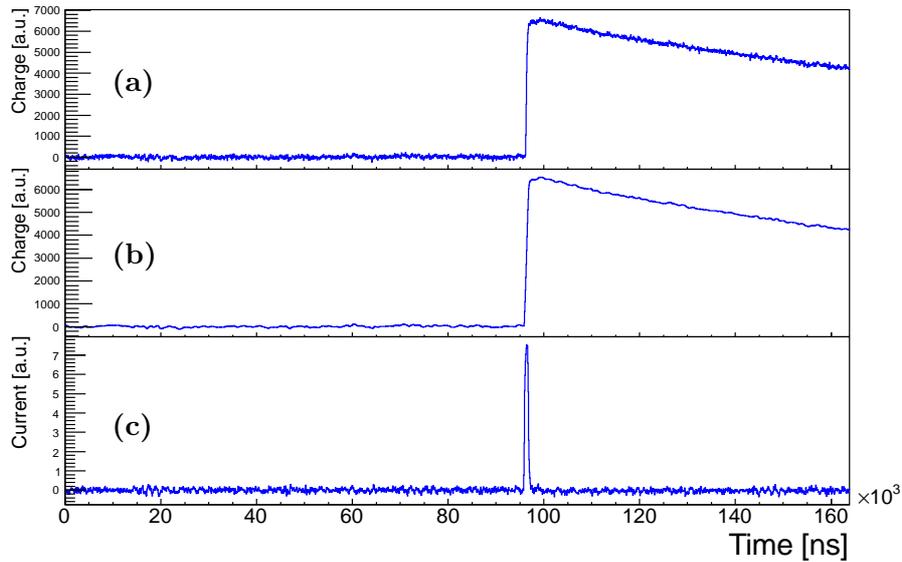,width=0.75\textwidth}
  \put(-300,180){\makebox(0,0)[l]{\textbf{(a)}}}
  \put(-300,115){\makebox(0,0)[l]{\textbf{(b)}}}
  \put(-300,50){\makebox(0,0)[l]{\textbf{(c)}}}
 
  \caption{An example of a transform chain: (a) Original waveform (charge pulse); (b) Waveform processed 
with a 200-ns moving average filter to reduce the noise; (c) Output of the 
MGWFDerivative transform applied to the waveform (b), which yields the current pulse.}
  \label{fig:TransformChain}
\end{center}
\end{figure}

A complex digital analysis of a real-life experiment can be implemented as a chain of \MGDO\ transforms, 
each performing a basic step of the processing. 
An example of chain, involving the smoothing and differentiation of a charge pulse, 
is depicted in Fig.~\ref{fig:TransformChain}.

\section{Real-life application of MGDO}
The tools available in the \MGDO\ library are used in the analysis software under development by \Majorana\ and 
\GERDA. 
Both the \GERDA\ analysis framework \textsc{Gelatio}~\cite{gel} as well as the \Majorana\ analysis toolkit ``GAT'' rely on \MGDO\ for the I/O structure and 
for the basic digital filters used in signal processing. Raw data produced in the \GERDA\ and \Majorana\ apparatuses as well as in 
detector test stands and simulations are transformed in \MGDO\ data objects and stored as 
\ROOT\ files. This allows the treatment of all data with the same analysis framework, irrespective of the native 
DAQ binary format and of the type of detector. Digitized traces from both HPGe detectors and 
photo-multipliers (e.g.~from veto detectors) are treated identically and are stored 
in arrays of MGWaveforms inside MGTEvent objects. Furthermore, the \textsc{Gelatio} and GAT analysis modules 
implementing the waveform analysis are built as chains of \MGDO\ transforms, thus favoring the 
re-use of the code and avoiding unnecessary duplication. 

Although the specific applications by the \Majorana\ and \GERDA\ experiments drove the design and the 
development of the toolkit, the \MGDO\ libraries, and in particular MGWaveform and its transforms, 
are generic enough to be applicable beyond the context of Germanium-based neutrinoless $\beta\beta$ 
decay experiments. It could be of potential interest for other 
low-background experiments from the perspective of setting up a standard and portable format which is 
suitable for the sharing and the inter-exchange of data and information.

\section*{Acknowledgments}
We would like to express our gratitude to the \Majorana\ and \GERDA\ Collaborations for 
the support given to this joint initiative.
We thank in particular R.~Henning for his continuous help 
and advice in the design and in the development of \MGDO.

\smallskip

\end{document}